\documentclass[runningheads]{llncs}
\usepackage{overpic}
\usepackage{amsmath}
\usepackage{amssymb}
\usepackage{graphicx}
\usepackage{enumerate}
\usepackage{enumitem}
\usepackage{booktabs}
\usepackage{bbding}
\usepackage{pifont}
\usepackage{wasysym}
\usepackage{amssymb}
\newcommand{\eg}[1]{\textit{e.g.,}}
\newcommand{\ie}[1]{\textit{i.e.,}}
\newcommand{\etc}[1]{\textit{etc}}
\usepackage{threeparttable}
\usepackage{color}
\usepackage{multirow}
\newcommand{\figref}[1]{Fig.\!~\ref{#1}}
\usepackage{colortbl}
\usepackage{bm}
\usepackage{mathtools}
\usepackage{siunitx}
\usepackage[nopar]{lipsum}
\usepackage{listings}
\usepackage[export]{adjustbox}
\usepackage{verbatim}
\usepackage{colortbl}
\usepackage[dvipsnames]{xcolor}
\newcommand{\stdvu}[1]{\scriptsize{\color{darkgray}(#1)} {\color{ForestGreen}$\uparrow$}}
\newcommand{\stdvd}[1]{\scriptsize{\color{darkgray}(#1)} {\color{red}$\downarrow$}}
\newcommand{\stdvw}[1]{\scriptsize{\color{darkgray}(#1)} {\color{white}$\downarrow$}}
\newcommand{\stdvno}[1]{\scriptsize{\color{darkgray}(#1)} {\color{mypink}$\downarrow$}}
\definecolor{mypink}{rgb}{.99,.91,.95}

\usepackage{etoolbox}
\makeatletter
\let\@algcomment\relax
\newcommand\algcomment[1]{\def\@algcomment{\footnotesize#1}}
\makeatother
\usepackage[pagebackref=false,breaklinks=true,colorlinks=true,bookmarks=false,citecolor=blue]{hyperref}
\begin{document}
\title{Enhancing Label-efficient Medical Image Segmentation with Text-guided Diffusion Models}  
% {\thanks{Supported by organization x.}}
% \author{Chun-Mei Feng\inst{1,2}
% \and
% Yong Xu \inst{1}}
% \institute{$^1$ Shenzhen Key Laboratory of Visual Object Detection and Recognition, Harbin Institute of Technology, shenzhen\\
% $^2$ Inception Institute of Artificial Intelligence \\
% {\tt \{strawberry.feng0304\}@gmail.com}\\
% {\tt \href{https://github.com/chunmeifeng/MINet}{https://github.com/chunmeifeng/MINet}}
% \thanks{This work was done during the internship of C.-M.~Feng at Inception Institute of Artificial Intelligence. Yong Xu is the corresponding author.}
% \author{Chun-Mei Feng$^1$}
% \institute{$^1$ Institute of High Performance Computing (IHPC),\\
% Agency for Science, Technology and Research (A*STAR)
% \\ \email{fengcm.ai@gmail.com}}
% {\tt \{fengcm.ai\}@gmail.com}\\
% {\tt \href{https://github.com/chunmeifeng/T2Net}{https://github.com/chunmeifeng/T2Net}}

\author{Chun-Mei Feng\inst{1}}
% % %
% % \authorrunning{F. Author et al.}
% % % First names are abbreviated in the running head.
% % % If there are more than two authors, 'et al.' is used.
% % %
\institute{$^1$ Institute of High Performance Computing (IHPC),\\
Agency for Science, Technology and Research (A*STAR)\\
{\tt fengcm.ai@gmail.com}\\
{\tt \href{https://github.com/chunmeifeng/TextDiff}{https://github.com/chunmeifeng/TextDiff}}}
\maketitle              % typeset the header of the contribution
\begin{abstract}
Aside from offering state-of-the-art performance in medical image generation, denoising diffusion probabilistic models (DPM) can also serve as a representation learner to capture semantic information and potentially be used as an image representation for downstream tasks, \eg, segmentation. However, these latent semantic representations rely heavily on labor-intensive pixel-level annotations as supervision, limiting the usability of DPM in medical image segmentation. To address this limitation, we propose an enhanced diffusion segmentation model, called TextDiff, that improves semantic representation through inexpensive medical text annotations, thereby explicitly establishing semantic representation and language correspondence for diffusion models. Concretely, TextDiff extracts intermediate activations of the Markov step of the reverse diffusion process in a pretrained diffusion model on large-scale natural images and learns additional expert knowledge by combining them with complementary and readily available diagnostic text information. TextDiff freezes the dual-branch multi-modal structure and mines the latent alignment of semantic features in diffusion models with diagnostic descriptions by only training the cross-attention mechanism and pixel classifier, making it possible to enhance semantic representation with inexpensive text. Extensive experiments on public QaTa-COVID19 and MoNuSeg datasets show that our TextDiff is significantly superior to the state-of-the-art multi-modal segmentation methods with only a few training samples. 
% \textit{Our code and models will be publicly available.}

\end{abstract}

\section{Introduction}
% Medical image segmentation aims to identify pixels of organs or lesions from background medical images, \eg, CT, X-Ray, or MRI images, to help doctors implement particular diagnoses and treatment plans~\cite{heller2021state,lenchik2019automated,rahman2020reliable,zhang2018task}. The rapid development of deep learning provides more development space for computer-aided diagnosis, which greatly reduces complexity compared with traditional mathematics-based methods~\cite{asiri2019deep}. For example, we can train a deep neural network through large-scale data so that the model can automatically identify and segment structures belonging to a certain disease tissue~\cite{wang2017two}.

The denoising Diffusion Probability Model (DPM) has recently demonstrated state-of-the-art performance in medical image generation~\cite{ho2020denoising,kazerouni2022diffusion,ruiz2022dreambooth,feng2023diverse,croitoru2022diffusion,kazerouni2022diffusion}, \eg, synthesis of pathological images~\cite{moghadam2023morphology}, generation of 3D brain MRI~\cite{peng2022generating,pinaya2022brain，feng2022multimodal}, and dynamic disease progression fitting~\cite{kim2022diffusion}, even surpassing GAN-based approaches.

% For example, Moghadam \textit{et al.} use the DPM to synthesize high quality histopathology images of brain cancer~\cite{moghadam2023morphology}. Based on DPM, Peng \textit{et al.} design an efficient strategy that generates high-fidelity 3D brain MRI~\cite{peng2022generating}. Pinaya \textit{et al.} use DPM to generate synthetic images from high-resolution brain images to complement training datasets~\cite{pinaya2022brain}. To dynamically capture the disease progression, Kim \textit{et al.} used the DPM to generate intermediate temporal volumes between source and target volumes to dynamically capture disease progression~\cite{kim2022diffusion}. 

Interestingly, recent work has found the potential of DPM as a representation learner to capture semantic information, as well as its advantages for downstream tasks such as natural image segmentation~\cite{baranchuk2021label}. 
%However, due to the difficulty in obtaining high-quality medical images required for lesion segmentation, as well as the labor-intensive pixel-level labeling of these images, the performance of deep learning-based medical image segmentation models, including the DPM, is severely limited~\cite{xu2019camel}. 
However, obtaining high-quality medical images necessary for lesion segmentation is difficult, and their pixel-level labeling is labor-intensive. 
As a result, the performance of deep learning-based medical image segmentation models, including DPM, is significantly limited~\cite{xu2019camel}.
%Given this, researchers are beginning to realize that over-reliance on labor-intensive pixel-level annotations as supervision to mine latent semantic representations is undesirable. 
This trend highlights the bottleneck caused by an over-reliance on labor-intensive pixel-level annotations as supervision to mine latent semantic representations.
%Intuitively, simply applying techniques such as semi-supervised learning~\cite{yu2019uncertainty,li2021dual} and weakly supervised learning~\cite{feng2017discriminative} to reduce the dependence of the deep model on the large amounts of annotated data appears to be effective. 
Instead, techniques like semi-supervised learning~\cite{yu2019uncertainty,li2021dual} and weakly supervised learning~\cite{feng2017discriminative} are being applied to reduce the deep model's dependence on large amounts of annotated data. 
%Disappointingly, the performance of such techniques heavily depends on the confidence of the pseudo-labels, and that a large number of pseudo-labels with low confidence will significantly lower the segmentation accuracy~\cite{li2022lvit}, thereby greatly reducing the clinical applicability of deep learning techniques. 
Unfortunately, the effectiveness of these techniques heavily relies on the confidence of the pseudo-labels. If a large number of pseudo-labels have low confidence, the segmentation accuracy can be significantly hampered~\cite{li2022lvit}, which greatly limits the clinical applicability of deep learning techniques. 
\textit{Therefore, how to develop an effective label-efficient diffusion model for medical image segmentation remains an unresolved question.}

As a remedy, we seek to increase the data usability by extracting knowledge from other 
%information that is complementary and readily available to medical images, \eg, medical text diagnostic information.
readily available sources of medical information, such as text diagnostic information, to complement medical images.
Medical text records are usually generated alongside sampled images, and accessing text diagnostic information corresponding to the images incurs no additional cost \cite{li2022lvit}.
%In general, the text medical records are usually generated along with the sampled images, thereby there is no additional cost to access the text diagnostic information corresponding to the images~\cite{li2022lvit}. 
The text diagnostic information records additional information complementary to image data. 
Huang \textit{et al.} leverage the radiology reports to learn global and local representations by contrasting image sub-regions and text annotations~\cite{huang2021gloria}, while Li \textit{et al.} introduce the medical text annotation to compensate the vision transformer~\cite{li2022lvit}. 
These methods demonstrate the usefulness of text diagnosis in image diagnosis using deep learning technology.  
\textit{Despite recent progress, it is unclear whether medical text diagnosis can also benefit the performance of diffusion models on medical image segmentation.} 
%With this perspective, we explore how to directly address the above issues by using additional medical text diagnostic information with the diffusion model~\cite{huang2021gloria}.
Hence, we investigate how additional medical text diagnostic information can directly address the aforementioned issues by incorporating it into the diffusion model~\cite{huang2021gloria}.

In this paper, we improve the performance of diffusion models in medical segmentation from the perspective of mining inexpensive medical text diagnostic information, yielding a new algorithm TextDiff that exhibits strong performance compared to various state-of-the-art multi-modal segmentation algorithms. Our main contributions are as follows:

% In this paper, we propose an enhanced label-efficient medical image segmentation method with text-guided diffusion models, termed TextDiff, making the diffusion model capture high-level semantic information valuable for downstream vision tasks in a small number of training images. TextDiff improves the visual semantic representation in the diffusion model by learning additional expert knowledge through inexpensive medical text annotation, enabling the model to perform well on a small number of training images. To further establish a strong connection between text diagnostic information and visuals, we design a cross-modal attention mechanism to match the text diagnostic descriptions with visual signals. Our main contributions are as follows: 
\vspace{-5pt}
\begin{enumerate}[leftmargin=20px]
    \item We propose an enhanced label-efficient medical image segmentation method, termed TextDiff, to reduce the dependence of the diffusion model on pixel-level annotations by learning additional expert knowledge through medical text annotations.
    
    % We propose an enhanced label-efficient medical image segmentation method with text-guided diffusion models to learn additional expert knowledge through cheap medical text annotations, enabling the diffusion model to capture high-level semantic information that is valuable for downstream vision tasks in a small number of training images.
    
    \item We establish strong connections between \textit{textual diagnostic annotations} and \textit{intermediate activations} of the Markov step of the reverse diffusion process in DPM, thereby improving visual-semantic representations in diffusion models. 
    % and thereby explicitly establishing semantic representations and linguistic correspondences for diffusion models.
    
    % We establish a strong connection between text diagnostic information and images via a cross-modal attention mechanism, thereby enhancing semantic representations.
    
    \item We \textit{freeze} the two-branch structure of TextDiff while \textit{only training} the cross-attention and pixel classifier, yielding significantly better results than various state-of-the-art multi-modal segmentation methods on COVID and pathological images with very few training samples, \eg, compared with GLoRIA~\cite{huang2021gloria}, TextDiff obtain the results of Dice: $66.38\%$ $\rightarrow$ $\textbf{78.67\%}$ and IoU: $49.83\%$ $\rightarrow$ $\textbf{64.98\%}$ on MoNuSeg dataset.
    
    % We perform extensive segmentation experiments on COVID and pathology images, demonstrating that our TextDiff significantly outperforms various state-of-the-art multi-modal segmentation methods with only a small number of training samples.
\end{enumerate}

\section{Methodology}
\subsection{Overall Architecture}
Given a medical image to be segmented, our goal is to train a deep neural network to automatically localize the visual region spatially of a certain tissue or lesion that the doctors are interested in. Here, unlike the previous works on diffusion models in image generation, we further explore the ability of DPM to capture high-level semantic information. Existing works train the network using visual information~\cite{baranchuk2021label}; on the contrary, we enhance the visual-semantic information by introducing inexpensive text diagnostic annotations that provide more efficient results. Such mechanism reduces the segmentation model's reliance on pixel-level annotations.

\begin{figure}[!t]
\centering
  \includegraphics[width=1\textwidth]{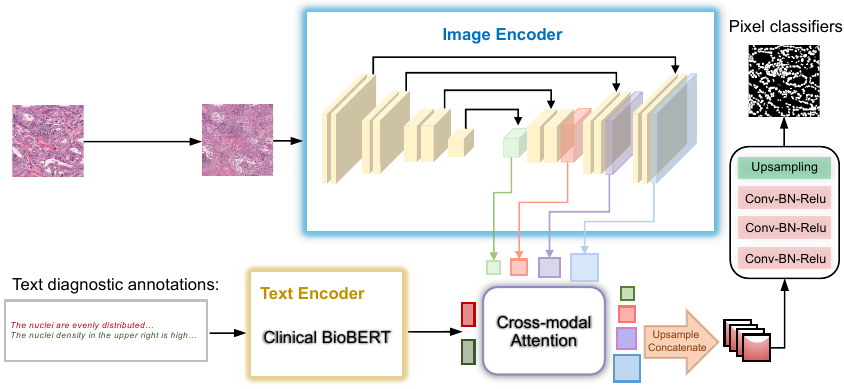}
  \put(-332,120){$\mathbf{x}_0$}
  % \put(-315,95){\tiny\rotatebox{90}{$q\!\left(\!\mathbf{x}^i_t \!\!\mid \!\mathbf{x}^i_{\!t\!-\!1}\!\right)$}}
  \put(-309,108){\small{$q\left(\mathbf{x}_t \!\mid \!\mathbf{x}_{t\!-\!1}\right)$}}
  \put(-253,120){$\mathbf{x}_t$}
  \put(-214,77){\small$\epsilon_\theta\left(\mathbf{x}_t, t\right)$}
  \put(-8,126){$\hat{\mathbf{x}}$}
  \put(-113,143){{\color{cyan}$\mathcal{E}_\texttt{{Im}}$}}
  \put(-194,37){{\color{brown}$\mathcal{E}_\texttt{{Te}}$}}
  \put(-256,28){$\mathbf{t}$}
  \put(-172,28){$\hat{\mathbf{t}}$}
  \put(-150,87){$\mathbf{h}_{0}$}
  \put(-127,82){$\mathbf{h}_{1}$}
  \put(-110,79){$\mathbf{h}_{2}$}
  \put(-92,75){$\mathbf{h}_{z}$}
  \put(-22,28){$\mathcal{H}$}
  % \put(-132,72){$\mathbf{h}$}
  % \put(-112,74){$\mathbf{h}$}
  % \put(-95,79){$\mathbf{h}$}
  % \vspace{-2pt}
  \caption{\textbf{Overview of the proposed TexDiff framework}, where the \textit{Image Encoder} is based on a pre-trained \texttt{Diffusion}~\cite{dhariwal2021diffusion} model to produce the high-level semantic information, while \texttt{Clinical BioBERT}~\cite{alsentzer2019publicly} serves as the \textit{Text Encoder}. \textit{Multi-scale Cross-modal Attention} leverages the knowledge of the text diagnostic annotation and images to be aligned for enhancing semantic representations.} 
  \label{fig1} 
  \vspace{-10pt}
\end{figure}
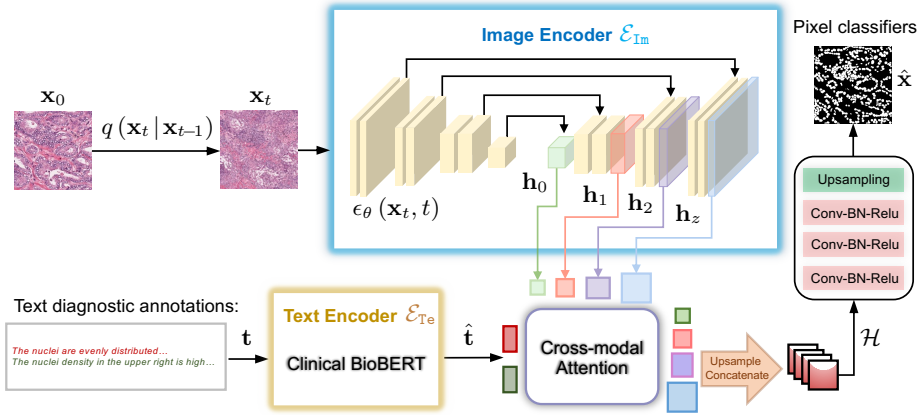

Since these texts are generated simultaneously with the diagnostic images, our training samples can be expressed as $\mathcal{D} = \left\{\left(\mathbf{x}^1,\mathbf{t}^1\right),\left(\mathbf{x}^2,\mathbf{t}^2\right),\ldots,\left(\mathbf{x}^N,\mathbf{t}^N\right)\right\}$, where $\mathbf{x}$, $\mathbf{t}$ refer to the diagnostic images (\eg, CT, X-Ray, or MRI images) and their corresponding text diagnostic annotation, respectively. As shown in Fig.~\ref{fig1}, the proposed TextDiff extracts visual and textual information by vision-language dual-branch architecture, \ie, diffusion model with UNet architecture~\cite{dhariwal2021diffusion} and \texttt{Clinical BioBERT}~\cite{alsentzer2019publicly}, respectively, and finally establish connections between textual diagnostic information and intermediate activations of the Markov step of the reverse diffusion process in DPM. Specifically, our TextDiff accepts two different modalities as input, \ie, $\mathbf{x}$ and $\mathbf{t}$. Each modality is sent to the pre-trained model to obtain the multi-modal feature representations, which are fused and then used to obtain predicted segmentation results by the pixel classifier. Note that only the multi-scale cross-attention and the pixel classifier are trainable in our method. We fix the weights of both the text encoder and the image encoder to maintain the vision-language alignment. To demonstrate the effectiveness of our proposed method, here, we simply use the means dice loss $\mathcal{L}_\texttt{Dice}$ and cross-entropy loss $\mathcal{L}_\texttt{CE}$ to evaluate the segmentation performance.

  % including an \textit{Image Encoder}, a \textit{Text Encoder}, and a \textit{Multi-scale Cross-modal Attention}. 

\vspace{2pt}
% \subsubsection{Image Encoding.} \label{sec:x1}
\noindent{\textbf{Image Encoding.}}
For a scanned image $\mathbf{x}$, we can obtain the features by a visual backbone $\hat{\mathbf{x}}=\mathcal{E}_\texttt{{Im}}(\mathbf{x})$. In our method, we explore whether the diffusion models can serve as an powerful instrument for segmentation. In image generation, diffusion models are used to transform noise $\mathbf{x}_T \sim N(0, I)$ to the sample $\mathbf{x}_0$ by gradually denoising $\mathbf{x}_T$ to less noisy samples $\mathbf{x}_t$. Formally, the forward diffusion process can be expressed as:
\begin{equation}
q\left(\mathbf{x}_t \mid \mathbf{x}_{t-1}\right):=\mathcal{N}\left(\mathbf{x}_t ; \sqrt{1-\beta_t} \mathbf{x}_{t-1}, \beta_t I\right).
\end{equation}
where $\beta_1, \ldots, \beta_t$ are the fixed variance schedule. 

Mathematically, the pre-trained DPM approximates a reverse process which can be expressed as follows
\begin{equation}
p_\theta\left(\mathbf{x}_{t-1} \mid \mathbf{x}_t\right):=\mathcal{N}\left(\mathbf{x}_{t-1} ; \mu_\theta\left(\mathbf{x}_t, t\right), \Sigma_\theta\left(\mathbf{x}_t, t\right)\right).
\end{equation}
Here, for an image input $\mathbf{x}\in \mathbb{R}^{H \times W \times C}$, we can compute $T$ sets of activation tensors from the noise predictor $\epsilon_\theta\left(\mathbf{x}_t, t\right)$ which is typically parameterized by different variants of the UNet architecture~\cite{dhariwal2021diffusion}. As shown in Fig.~\ref{fig1}, we first add Gaussian noise to corrupt $\mathbf{x}_0$. The parameterization of the UNet model $\epsilon_\theta\left(\mathbf{x}_t, t\right)$ uses the noisy $\mathbf{x}_t$ as an input. Bilinear interpolation is then used to upsample the intermediate activations of the UNet to $H \times W$. Such mechanism enables them to be treated as pixel-level representations of $\mathbf{x}_0$. 

Based on this, we further extract the pixel-level representation of the labeled image through UNet blocks and diffusion steps $t$, \eg, the middle block $B=\{4,6,7,8,12,16\}$ of UNet decoder and steps $t=\{50,150,250\}$ are adopt for the feature extraction~\cite{baranchuk2021label}. Then, to create feature vectors $\hat{\mathbf{x}}$ for all of the pixels in the training images, the extracted representations from these blocks and steps are fused with medical text features $\hat{\mathbf{t}}$ produced by text encoder $\mathcal{E}_\texttt{{Te}}$. Note that we only consider decoder activations because the skip connections also collect encoder activations. Finally, we can obtain the prediction results $\hat{\mathbf{x}}$ for each pixel by training the pixel classifier while freezing the two-branch backbone.

\vspace{2pt}
\noindent{\textbf{Text Encoding.}}\label{sec:x2}
% \subsubsection{Text Encoding.} 
As we mentioned before, the textual diagnostic annotations are generated without extra collection cost alongside the sampled images, and their small footprint makes them a natural complement to image data. As a result, as shown in Fig.~\ref{fig1}, we use a pre-trained text encoder, \ie, \texttt{Clinical BioBERT}~\cite{alsentzer2019publicly}, to extract valuable information from the text diagnostic annotations. \texttt{Clinical BioBERT}~\cite{alsentzer2019publicly} is the pre-trained text model which obtain the clinical-aware text embeddings on the MIMIC III dataset~\cite{johnson2016mimic}. Specifically, given a annotations ${\mathbf{t}}$, we can obtain the features by a text backbone $\hat{\mathbf{t}}=\mathcal{E}_\texttt{{Te}}(\mathbf{t})$. Subsequently, the text features are fed into the cross-modal attention along with the intermediate activations of the Markov step of the reverse diffusion process in DPM. Details are provided in Sec.~\ref{sec:x4}.

% \vspace{5pt}
% \noindent{\textbf{Loss Function.}} \label{sec:x3}
% To demonstrate the effectiveness of our proposed method, here, we simply use the means dice loss $\mathcal{L}_\texttt{Dice}$ and cross-entropy loss $\mathcal{L}_\texttt{CE}$ to evaluate the segmentation performance. Formally, the loss function can be expressed as $\mathcal{L}_\texttt{} = \mathcal{L}_\texttt{Dice}+\mathcal{L}_\texttt{CE}$. 

% Given that model distillation can benefit two different networks from each other, we use it to further enhance the interaction of text and visual information. To do this, we first align these features before projecting them into the latent space. Specially, $\mathcal{H}$ and $\hat{\mathbf{t}}$ are first normalized by \texttt{softmax} and then learn the mutual information by the following loss:
% \begin{equation}
% \mathcal{L}_{KL}=D_{K L}\left(\operatorname{softmax}\left(\mathcal{H}\right) \|\left(\operatorname{softmax}\left(\hat{\mathbf{t}}\right)\right)\right.,
% \end{equation}
% where $D_{KL}$ is the Kullbach Leibler (KL) Divergence. Additionally, we use the means dice loss $\mathcal{L}_\texttt{Dice}$  and cross-entropy loss $\mathcal{L}_\texttt{CE}$ to evaluate the segmentation performance. Formally, our overall loss function can be expressed as $\mathcal{L} = \alpha \mathcal{L}_{\texttt{KL}}+(1-\alpha)(\left\mathcal{L}_\texttt{Dice}+\mathcal{L}_\texttt{CE}\right)$, 
% where $\alpha$ weights the trade-off between the different losses.

% 
\subsection{Cross-modal Attention for Knowledge Alignment} \label{sec:x4}
Here, we consider how to align the text and visual features, thereby enhancing the visual semantic representation. We define a cross-modal attention module $\mathcal{M}_{\texttt{cro}}$ that integrates features of different sizes in the diffusion model of an UNet shape, thereby cross-contextualizing the text embeddings with pixel representations of the image. The cross-modal attention module provides a strong connection between language and vision, enabling textual information to enhance semantic representation in images~\cite{feng2021task}. 
% Our cross-modal attention module takes the text information $\hat{\mathbf{t}}$ extracted from the text encoder and the pixel-level visual features ${\mathbf{h}}_z$ with different sizes extracted from the diffusion model as input. 
% Formally, we have
% Following~\cite{tan2019lxmert}, we treat the visual as query ($Q$) and text feature as key ($K$) and value ($V$), separately. The idea is to query the corresponding visual features in the image through language instructions, thereby ensuring that the corresponding text vocabulary matches the visual features and enhancing the representation of semantic features. 
Specifically, given the pixel-level visual features, ${\mathbf{h}}_z$, where $i=0,1,...,z$, with different sizes extracted from the diffusion decoder and its corresponding text feature $\hat{\mathbf{t}}$, we compute the scaled dot-product attention at the step $t$:
\begin{equation}
\begin{gathered}
{\mathcal{H}_{z,t}}=\operatorname{Softmax}\left({\mathbf{h}}_{z,t} W_q ({\hat{\mathbf{t}}}_{} W_k)^T/{\sqrt{d}}\right) \hat{\mathbf{t}} W_v,
\end{gathered}
\end{equation}
where $W_q, W_k$, and $W_v$ are the learned parameter matrices. ${\mathcal{H}}_{z,t}$ is the different scales attention representation over the text enhanced visual medical image. We concatenate these attention representations to get ${\mathcal{H}}_{}$ after upsampling them to the same size. We analyze whether medical text diagnosis brings benefits to the performance of diffusion models on medical image segmentation in Sec.~\ref{ab}.

% We note that the self-attention is first applied to text and image features separately followed by the cross-modal attention~\cite{vaswani2017attention}.

\section{Experiments} \label{sec:result}
% \subsubsection{Experimental Setup.}
% \vspace{5pt}
\noindent{\textbf{Experimental Setup.}}~We train our method by Pytorch with one NVIDIA Tesla V$100$ GPU and $32$GB of memory. We use Adam as the optimizer, with an initial learning rate of $1$e$-4$ and a batch size of $1$, for $100$ epochs. The input images are resized to
$256\times256$. The middle blocks $B=\{6, 8, 12, 16\}$ and $B=\{4, 6, 8, 12\}$ of the UNet decoder with steps $t=\{50,150,250\}$ are adopted for MoNuSeg and QaTa-COVID19, respectively.

\begin{table*}[t]
\renewcommand{\arraystretch}{1.3}
	\caption{\small \textbf{Quantitative comparison} of state-of-the-art methods on two datasets, where {\# Param} is the parameter cost,  ${\color{ForestGreen}\uparrow}$ and ${\color{red}\downarrow}$ indicate increments and decrements compared with UNet, respectively. Detailed analyses are provided in Sec.\ref{sota}. }
	% \vspace{-5pt}
	\label{tab:1}

	\fontsize{9}{9}\selectfont
	\centering
	\begin{tabular}{l c cc cc cc cc}
\toprule
\multicolumn{1}{l}{}&&\multicolumn{4}{c}{{{MoNuSeg}}}&\multicolumn{4}{c}{{{QaTa-COVID19}}} \\%\midrule
\cmidrule(lr){3-6} \cmidrule(l){7-10}
{Method}&\multicolumn{1}{c}{{~\# Param.~}}&\multicolumn{2}{c}{{~~~~Dice (\%)~~~~}} &\multicolumn{2}{c}{{~~~~IoU (\%)~~~~}} &\multicolumn{2}{c}{{~~~~Dice (\%)~~~~}} &\multicolumn{2}{c}{{~~~~IoU (\%)~~~~}} \\

\cmidrule(r){1-1} \cmidrule{2-2} \cmidrule(lr){3-4} \cmidrule(lr){5-6} \cmidrule(lr){7-8} \cmidrule(lr){9-10}  
UNet${\color{magenta}_{2015}}$\cite{ronneberger2015u} &\multicolumn{1}{|c|}{$31.04$ M}  &\multicolumn{2}{c}{$73.92$\stdvw{$00.00$}} &\multicolumn{2}{c}{$58.98$\stdvw{$00.00$}}&\multicolumn{2}{|c}{$46.08$\stdvw{$00.00$}} &\multicolumn{2}{c}{$34.16$\stdvw{$00.00$}} \\
TransUNet${\color{magenta}_{2021}}$\cite{chen2021transunet}&\multicolumn{1}{|c|}{$93.19$ M}  &\multicolumn{2}{c}{$73.54$\stdvd{$00.38$}} &\multicolumn{2}{c}{$58.79$\stdvd{$00.19$}}&\multicolumn{2}{|c}{$70.78$\stdvu{$24.70$}} &\multicolumn{2}{c}{$59.50$\stdvu{$25.34$}} \\
SwinUNet${\color{magenta}_{2021}}$\cite{cao2021swin} &\multicolumn{1}{|c|}{$27.17$ M}  &\multicolumn{2}{c}{$64.36$\stdvd{$09.56$}} &\multicolumn{2}{c}{$48.74$\stdvd{$10.24$}}&\multicolumn{2}{|c}{$65.19$\stdvu{$19.11$}} &\multicolumn{2}{c}{$51.87$\stdvu{$17.71$}} \\
% MedT${\color{magenta}_{2021}}$\cite{valanarasu2021medical} &\multicolumn{1}{|c|}{$1.56$ M}  &\multicolumn{2}{c}{0\stdvd{$0$}} &\multicolumn{2}{c}{0\stdvd{$0$}}&\multicolumn{2}{|c}{0\stdvd{$0$}} &\multicolumn{2}{c}{0\stdvd{$0$}} \\
GLoRIA${\color{magenta}_{2021}}$\cite{huang2021gloria} &\multicolumn{1}{|c|}{$32.52$ M}  &\multicolumn{2}{c}{$66.38$\stdvd{$07.54$}} &\multicolumn{2}{c}{$49.83$\stdvd{$09.15$}}&\multicolumn{2}{|c}{$71.05$\stdvu{$24.97$}} &\multicolumn{2}{c}{\textbf{59.74}\stdvu{$24.58$}} \\
LViT${\color{magenta}_{2022}}$\cite{li2022lvit} &\multicolumn{1}{|c|}{$29.72$ M}  &\multicolumn{2}{c}{$57.95$\stdvd{$15.97$}} &\multicolumn{2}{c}{$44.13$\stdvd{$14.85$}}&\multicolumn{2}{|c}{$66.43$\stdvu{$20.35$}} &\multicolumn{2}{c}{$51.71$\stdvu{$17.55$}} \\
% MedKLIP${\color{magenta}_{2023}}$\cite{wu2023medklip} &\multicolumn{1}{|c|}{$0$ M}  &\multicolumn{2}{c}{0\stdvd{$0$}} &\multicolumn{2}{c}{0\stdvd{$0$}}&\multicolumn{2}{|c}{0\stdvd{$0$}} &\multicolumn{2}{c}{0\stdvd{$0$}} \\
{\cellcolor{mypink}$\textbf{{TextDiff}}$ \texttt{(Ours)}} &\multicolumn{1}{|c|}{{\cellcolor{mypink}$\textbf{9.68}$ M}} &\multicolumn{2}{c}{{\cellcolor{mypink}\textbf{78.67}\stdvu{$\underline{04.75}$}}} &\multicolumn{2}{c}{{\cellcolor{mypink}\textbf{64.98}\stdvu{$\underline{06.00}$}}}&\multicolumn{2}{|c}{{\cellcolor{mypink}\textbf{71.41}\stdvu{$\underline{25.33}$}}}&\multicolumn{2}{c}{{\cellcolor{mypink}\textbf{$59.03$}\stdvu{$\underline{24.87}$}}} \\
\bottomrule
\end{tabular}
% \vspace{-10pt}
\end{table*}

\vspace{3pt}
% \subsubsection{Datasets.}
\noindent{\textbf{Datasets.}}~We employ two public datasets to evaluate our method, \ie, \textbf{1)} \textbf{MoNuSeg}~\cite{kumar2019multi} is a pathology dataset obtained from the MICCAI $2018$ MoNuSeg challenge and consists of $30$ images with $21,623$ nuclear boundary annotations for training and $14$ images with $7000$ nuclear boundary annotations for testing. To demonstrate the effectiveness of our label-efficient segmentation mechanism, we experiment with only a small number of images to clearly demonstrate the advantages of our method. We randomly select $5$ images from \textbf{MoNuSeg} for \texttt{training}, while the test set remains unchanged;  \textbf{2)} \textbf{QaTa-COVID19}~\cite{degerli2022osegnet} is collected from Qatar University and Tampere University and consist of $9258$ COVID-19 chest radiographs with pixel annotations of COVID-19 lesions. In our experiments, we randomly select $150$ images for \texttt{training}. Following~\cite{li2022lvit}, we use their extended text annotations to enhance the vision-language model. 

\vspace{3pt}
\noindent{\textbf{Baselines.}}
We compare two categories of methods to demonstrate the effectiveness of our proposed method, including \textbf{a)} classical medical segmentation methods: (1) UNet~\cite{ronneberger2015u}, (2) TransUNet~\cite{chen2021transunet}, (3) SwinUNet\cite{cao2021swin}, and \textbf{b)} text-driven medical segmentation methods: (1) GLoRIA~\cite{huang2021gloria}, a multi-modal medical image recognition framework that learns the global and local representations by contrasting image sub-regions and text in the paired report; (2) LViT~\cite{li2022lvit}, a multi-modal medical image segmentation framework based on the transformer that uses the medical text annotations to compensate for the visual representation. For a fair comparison, we retrain all the baseline methods with their default parameter and report the best results.

\subsection{Comparison with State-of-the-arts.}\label{sota}
To investigate the effectiveness of our method, we show the comparison results, \ie, Dice (\%) and IoU (\%), with various state-of-the-art methods in Table~\ref{tab:1}. Our approach yields the highest values on all datasets with regard to Dice (\%) and IoU (\%). Specifically, the classical medical segmentation methods, \ie, UNet~\cite{ronneberger2015u}, and TransUNet~\cite{chen2021transunet}, SwinUNet\cite{cao2021swin} are less effective than the language-vision methods, \ie, GLoRIA~\cite{huang2021gloria}, LViT~\cite{li2022lvit}, and our proposed method. However, the segmentation results of multi-modal methods, \ie, GLoRIA~\cite{huang2021gloria} and LViT~\cite{li2022lvit}, are still lower than our method. This is mainly due to the deep alignment of text and visual information extracted by the diffusion model in our method. Although both the GLoRIA~\cite{huang2021gloria} and LViT~\cite{li2022lvit} absorbed the text information, LViT requires more parameters, \ie, $29.72$ M, and GLoRIA cannot provide effective visual features. Besides, LViT needs to be trained from scratch, while our model takes advantage of the powerful large-scale pre-training model on natural images. The dual-branch text and image encoders of our method are frozen, and we only need to update the pixel classifier and cross-attention mechanism. In particular, compared with the state-of-the-art vision-language medical method GLoRIA~\cite{huang2021gloria}, our method improves the Dice (\%) values from $66.38$ to {\textbf{78.67}}, and the IoU (\%) values from $49.83$ to $\textbf{64.98}$ on the MONuSeg dataset. More importantly, our proposed method only requires $9.68$ M of communication. These demonstrate that our model can effectively enhance the visual features through the text annotations, which is beneficial to medical image segmentation and reduces reliance on labor-intensive pixel-level annotation as supervision.

\begin{figure}[!t]
\centering
  \includegraphics[width=1\textwidth]{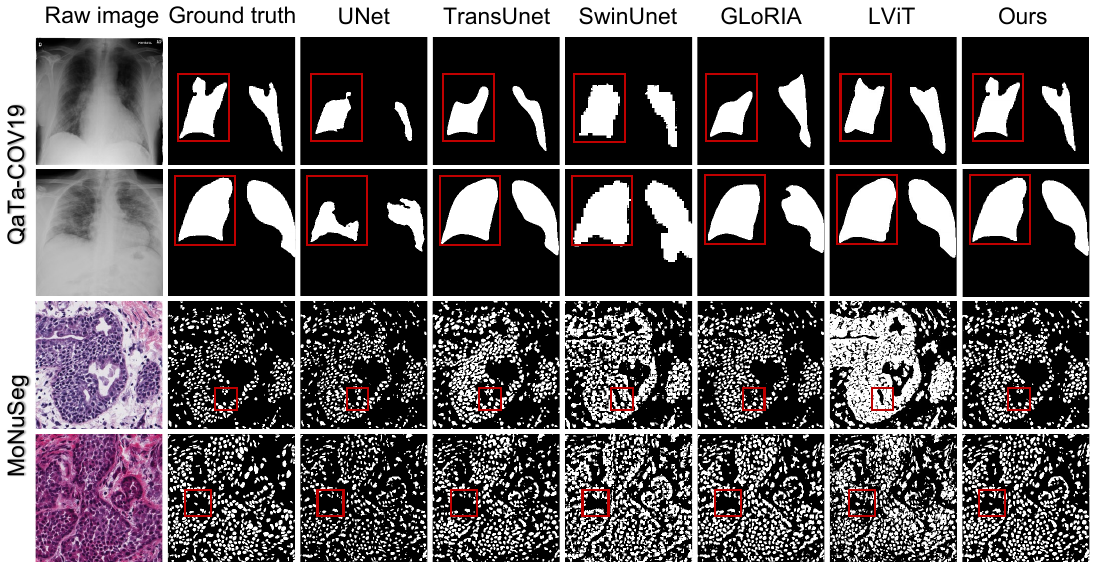}
  % \vspace{-15pt}
  \caption{Visual segmentation comparisons with regards to different datasets, detailed analysis is provided in Sec.~\ref{sota}.} 
  \vspace{-4pt}
  \label{figure2} 
\end{figure}

\figref{figure2} provides the qualitative results of our model and other state-of-the-art methods on the two datasets. As compared with the ground-truth, the language-vision methods outperform the classical medical segmentation methods. Notably, our method provides fewer errors and segmentation results that are closest to ground-truth, which is attributable to the fact that our method can effectively learn enhanced features from the inexpensive text annotations. Further, these visual segmentation results from different datasets, \ie, MoNuSeg~\cite{kumar2019multi} and QaTa-COVID19~\cite{degerli2022osegnet}, support our conclusion that our method can leverage the medical text diagnosis to benefit the performance of diffusion models on medical image segmentation.

\subsection{Ablation Studies.}\label{ab}

\begin{figure}[!t]
\centering
  \includegraphics[width=1\textwidth]{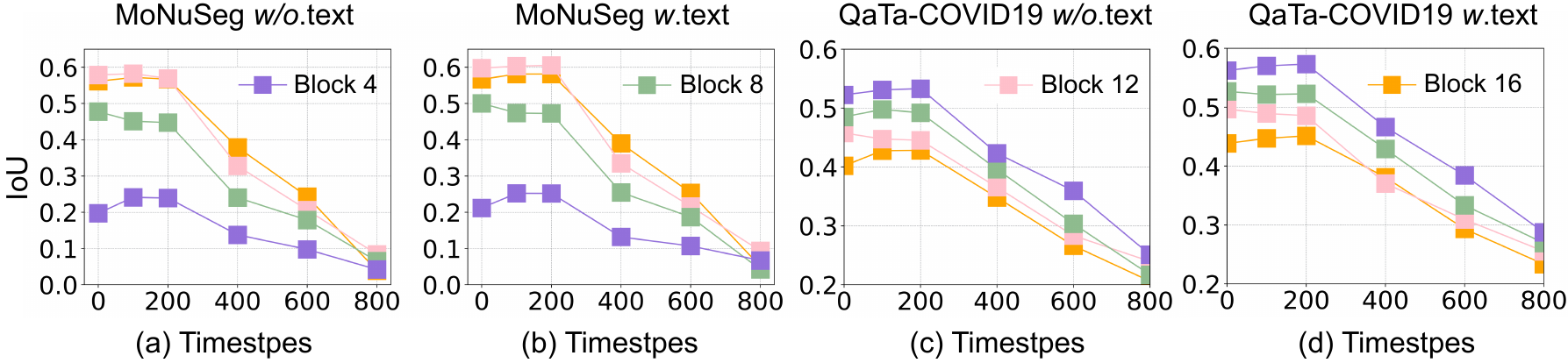}
  % \vspace{-20pt}
  \caption{Evolution of the segmentation performance with regard to different blocks and steps of our proposed method on the two datasets, see Sec.\ref{ab} for more details.}
  \label{figure3} 
\end{figure}

% \vspace{4pt}
\noindent{\textbf{Representation analysis.}}
To investigate the performance of medical text diagnostic annotations in the diffusion representations, we show the evolution of prediction performance over different blocks and diffusion steps $t$ in Fig.~\ref{figure3}, where (a) and (c) are the representations without text, (b) and (d) are the representations with text. As can be seen from this figure, the semantic representations produced by the diffusion model vary for different blocks and diffusion steps, and these representations are enhanced to varying degrees after introducing text diagnostic annotations (see Fig.~\ref{figure3} (b) and (c)). These experimental results provide an \textbf{\textit{answer}} to the question of whether medical text diagnosis can improve the performance of diffusion models on medical image segmentation. Additionally, we find that features corresponding to later steps in the reverse diffusion process are often more effective at capturing semantic information, while the ones corresponding to earlier steps are generally uninformative. In different blocks, the features produced by different layers of the UNet decoder on the two datasets seem to be different, allowing us to choose different blocks for different datasets in our experiments.

% \vspace{5pt}
\noindent{\textbf{Key components analysis.}}
Here, we evaluate the effectiveness of the key components of our method through some variations of it, \ie, $\zeta_1$, which is our method without medical text annotations, and $\zeta_2$, which is our method without multi-scale cross attention. We summarize the results of these ablation models in Table~\ref{tab:2}. As can be seen from this table, we observe that $\zeta_1$ performs the worst, which is consistent with our primary motivation that the text annotations can provide supplementary information for visual features. Since without multi-scale cross-attention module cannot learn the deep information of the two modalities, the results of $\zeta_2$ are not optimal. Yet, our full TextDiff further aligns the features of the different modalities, and yields the best results, demonstrating its powerful capability in medical image segmentation.

% To study the effect of trade-off weights between different losses, we report the segmentation results of our method on both the \textbf{QaTa-COVID19} and \textbf{MoNuSeg} datasets in Fig.~\ref{figxxx}. The weight of $\mathcal{L}_{KL}$ is determined by the value of $\alpha$, where the larger the value of $\alpha$, the greater the influence of text knowledge on visual semantic information. It can be seen from the figure that when $\alpha = 0.7$, TextDiff gets the highest segmentation result, \ie, Dice (\%)=xx, IoU (\%)=xx. This shows that knowledge distillation can further enhance the interaction between different modalities. The segmentation results will decline when the values of $\alpha$ is less than $0.3$, which may indicate that the supervision is insufficient in such case.

\begin{table*}[t]
\renewcommand{\arraystretch}{1.3}
	\caption{ Ablation studies on the MoNuSeg dataset, where ${\color{red}\downarrow}$ indicates decrements compared with our full model TextDiff. Detailed analyses are provided in Sec.\ref{ab}. }
	% \vspace{-7pt}
	\label{tab:2}

	\fontsize{9}{9}\selectfont
	\centering
	\begin{tabular}{c c  c cc cc cc cc cc}
\toprule
\multicolumn{1}{l}{}&&&\multicolumn{4}{c}{{{\textbf{MoNuSeg}}}}&\multicolumn{4}{c}{{{\textbf{QaTa-COVID19}}}}\\%\midrule
\cmidrule(lr){4-7} \cmidrule(lr){8-11} 
% \cmidrule(l){7-10}
~{Variation}~~&\multicolumn{1}{c}{~~{\texttt{Text}}~~}&\multicolumn{1}{c}{{~~$\mathcal{M}_{\texttt{cro}}$~~}}&\multicolumn{2}{c}{~~~~{Dice (\%)}~~~~} &\multicolumn{2}{c}{~~~~{IoU (\%)}~~~~} &\multicolumn{2}{c}{~~~~{Dice (\%)}~~~~} &\multicolumn{2}{c}{~~~~{IoU (\%)}~~~~} \\

\cmidrule(r){1-1} \cmidrule{2-2} \cmidrule(lr){3-3}  \cmidrule(lr){4-5}  \cmidrule(lr){6-7}  \cmidrule(lr){8-9}  \cmidrule(lr){10-11}

$\zeta_1$ &\multicolumn{1}{|c}{-}  &-  &\multicolumn{2}{|c}{77.73\stdvd{$0.73$}} &\multicolumn{2}{c|}{63.70\stdvd{$0.95$}} &\multicolumn{2}{c}{68.90\stdvd{$5.37$}} &\multicolumn{2}{c}{56.94\stdvd{$5.09$}}\\

$\zeta_2$ &\multicolumn{1}{|c}{\Checkmark}  &-  &\multicolumn{2}{|c}{76.34\stdvd{$2.12$}} &\multicolumn{2}{c|}{61.89\stdvd{$2.76$}} &\multicolumn{2}{c}{70.71\stdvd{$3.56$}} &\multicolumn{2}{c}{57.81\stdvd{$4.22$}}\\

\texttt{Ours} &\multicolumn{1}{|c}{\Checkmark}  &\multicolumn{1}{c}{\Checkmark}  &\multicolumn{2}{|c}{{\cellcolor{mypink}78.67\stdvno{$0.00$}}} &\multicolumn{2}{c|}{{\cellcolor{mypink}64.98\stdvno{$0.00$}}} &\multicolumn{2}{c}{{\cellcolor{mypink}71.41\stdvno{$0.00$}}} &\multicolumn{2}{c}{{\cellcolor{mypink}59.03\stdvno{$0.00$}}}\\

% LViT${\color{magenta}_{2022}}$\cite{li2022lvit} &\multicolumn{1}{|c|}{$0$ M}  &\multicolumn{2}{c}{0\stdvd{$0$}} &\multicolumn{2}{c}{0\stdvd{$0$}}&\multicolumn{2}{|c}{0\stdvd{$0$}} &\multicolumn{2}{c}{0\stdvd{$0$}} \\
% {\cellcolor{mypink}$\textbf{{TextDiff}}$ } &\multicolumn{1}{|c|}{{\cellcolor{mypink}$\textbf{0}$ M}} &\multicolumn{2}{c}{{\cellcolor{mypink}\textbf{0}\stdvu{$\underline{0}$}}} &\multicolumn{2}{c}{{\cellcolor{mypink}\textbf{0}\stdvu{$\underline{0}$}}}&\multicolumn{2}{|c}{{\cellcolor{mypink}\textbf{0}\stdvu{$\underline{0}$}}}&\multicolumn{2}{c}{{\cellcolor{mypink}\textbf{0}\stdvu{$\underline{0}$}}} \\
\bottomrule
\end{tabular}
% \vspace{-9pt}
\end{table*}

\section{Conclusion}
This work focuses on how to extend the diffusion model to the medical segmentation task with text annotations, thereby reducing the over-reliance on labor-intensive pixel-level annotations as supervision. Given this, we propose a diffusion segmentation method, termed TextDiff, that improves semantic representation via inexpensive medical text annotations, allowing the model to perform well on a small number of training images. To the best of our knowledge, TextDiff is the first multi-modal diffusion framework for medical image segmentation. 
% Specifically, our method consists of two encoders, \ie, image and text encoders, together with a multi-scale cross attention module to deeply extract and align the textual and visual knowledge for the pixel segmentation. 
Experiments on the different datasets demonstrate the superiority of TextDiff in medical image segmentation with limited training samples.

\bibliographystyle{splncs04}
\bibliography{bibliography}

\end{document}